\documentclass[%
 aip,
 jmp,%
 amsmath,amssymb,
 reprint,%
]{revtex4-1}

\usepackage{graphicx,color,epsfig,rotate}
\usepackage{dcolumn}
\usepackage{bm}

\begin{document}

\preprint{AIP/123-QED}

\title[]{Orientational correlations in fluids with quenched disorder}

\author{N. Shankaraiah}
 \email{naddis@tifrh.res.in}
 
\author{Surajit Sengupta}%
 \email{surajit@tifrh.res.in}
\affiliation{TIFR Centre for Interdisciplinary Sciences, 36/p Gopanpally, Hyderabad 500107, India}%

\author{Gautam I. Menon}
 \email{menon@imsc.res.in}
\affiliation{The Institute of Mathematical Sciences, C.I.T Campus, Taramani, Chennai 600113, India}%
\affiliation{ Homi Bhabha National Institute, Training School Complex, Anushaktinagar, Mumbai 400 094, India}%

\date{\today}

\begin{abstract}
Snapshots of colloidal particles moving on disordered two-dimensional substrates can be used to extract equal-time many-body correlations in their positions. To understand the systematics of these correlations, we perform Monte Carlo simulations of a two-dimensional model fluid placed in a quenched disordered background. We use configurations generated from these simulations to compute translational and orientational two-point correlations at equal time, concentrating on correlations in local orientational order as a function of density and disorder strength. We calculate both the disorder averaged version of conventional two-point correlation functions for orientational order, as well as the disorder averaged version of a novel correlation function of time-averaged disorder-induced inhomogeneities in local orientation analogous to the Edwards-Anderson correlation function in spin systems. We demonstrate that these correlations can exhibit interesting non-monotonic behaviour in proximity to the underlying fluid-solid transition and suggest that this prediction should be experimentally accessible.
\end{abstract}

\maketitle

\section{\label{sec:level1}Introduction}
Translational invariance is broken~\cite{CL} in systems with quenched disorder that is distributed inhomogeneously in space~\cite{shobo,gauflux,gauflux2}. Novel equal-time correlations~\cite{hansen}, otherwise trivial in the pure system, emerge as a result~\cite{gauchand}. These are experimentally measurable, provided one has access to snapshots of particle configurations in real space~\cite{egelme}. Such correlations can exhibit interesting behavior even in fully fluid phases~\cite{ankeur} of interacting particle systems, where issues of history dependence and the lack of equilibration can be ignored, in contrast to the glassy states in which they have traditionally been invoked and studied~\cite{gauchand}. 

Experiments on colloidal systems~\cite{sood1,sood2,hartmut1,hartmut2} are capable of probing  correlations of this form, since current imaging experiments allow us to access  the instantaneous positions of colloidal particles over a large field. Using colloids as model atomic systems has been particularly useful in furthering our understanding of non-equilibrium and equilibrium phases of matter~\cite{camporev}, including the structural glass problem~\cite{glassbook, glasscol}. Recent experiments on colloidal systems allow us to test ideas from theory~\cite{maret-mermin} and computer  simulations~\cite{bow,ghosh,zahn}. The response of colloids to spatially inhomogeneous external laser fields~\cite{jorg-rev} was first studied in the context of the freezing of colloidal liquid into a modulated crystal. Such freezing has been studied in experiments on a two-dimensional (2D) system of strongly interacting colloidal particles subjected to a one-dimensional periodic modulating potential~\cite{ack1,ack2,jay1}. Monte Carlo studies have obtained both laser-induced freezing as well as described the novel re-entrant melting transition  of a colloidal crystal to a modulated liquid~\cite{jay2}. This prediction has been experimentally verified~\cite{bechlif}. Experiments aimed at studying order disorder phenomena through direct visualization are possible in many other systems. These include colloids driven by magnetic ratchets~\cite{tierno}, magnetic bubble arrays~\cite{R1,R2} as well as a host of physical-chemical systems exhibiting domain patterns as a consequence of competing interactions~\cite{patt}. 

Given a large number of snapshots of the positions of colloidal particles moving on a substrate, one can construct arbitrary n-point correlations from them. Usually, the correlations of interest are one-body and two-body correlation functions. In pure fluid phases, the one-body densities are uniform~\cite{CL,hansen}, since translation invariance privileges no specific position over another. In addition, two-point correlations are functions only of the distance between the particles. In systems where the particles feel an underlying, typically quenched disordered, potential, translational invariance is broken and even one-body distributions become spatially inhomogeneous. Translational invariance is restored by the expedient of averaging over disorder realizations or simply by considering sufficiently large systems, assuming that self-averaging holds~\cite{ankeur}.

Some years ago, some of us showed that Monte Carlo simulations and liquid state theory approaches could be used to calculate two distinct types of  translational correlation functions in model two-dimensional colloidal systems in a quenched disordered background~\cite{gauchand,ankeur}. The diagonal correlation function is simply the disorder-averaged version of the standard pair correlation function in the pure system. The off-diagonal correlation function, on the other hand, represents the equal-time correlation of disorder-induced density inhomogeneities. It is the analog of the Edwards-Anderson order parameter (EA-OP)~\cite{EA}. Recent experiments on colloids in the presence of quenched disorder, realised by generating random potential energy landscapes using laser speckles,  measured these  correlation functions~\cite{egelme}, finding  very good agreement with the predictions of liquid-state theory~\cite{gauchand} and simulations~\cite{ankeur}. However, the question of how similarly defined correlation functions for orientational order might behave was not explored in our earlier work, although they can now be accessed using similar measurements.

In this paper, we present results on disorder-averaged two-point translational and orientational correlations obtained via Monte Carlo simulations of a model fluid in a quenched disorder background\cite{gauchand, ankeur}. We aim to provide a model description of correlation functions that can be explored in experiments, in particular those that describe two-point correlations in local orientation. We find that the correlations of time-averaged local orientational correlations, 
has an intermediate distance limit that increases with disorder in the liquid regime while it decreases  with increasing disorder strength deep into the solid regime.  In this intermediate separation regime, the behavior of this correlation function shows re-entrance close to the putative liquid-solid phase boundary. The first peak height of the structure factor with disorder is roughly constant in the liquid regime,  is observed to decrease in the solid regime, while remaining flat initially and then decreasing as  the phase boundary is approached from the liquid side.  We provide qualitative, physical arguments for 
 this behavior.

The rest of the paper is organized as follows. In the next section (Section~\ref{model}) we summarize the correlation functions studied in this work, describing in detail the model system used for this study. Section~\ref{results} presents our results. Last, in Section~\ref{conclude} we outline the conclusions of this study and suggest future directions for related research.

\section{Model and Method}
\label{model}

Our model Hamiltonian~\cite{ankeur} (apart from the trivial and irrelevant contribution from momenta)  is given by, 
\begin{equation}
 H_{int}=\epsilon \sum_{i<j} \left(\frac{\sigma_0}{r_{ij}}\right)^{12} + \sum_{i} V_{d} (r_i).
 \label{eq:hamiltonian}
\end{equation}
We set $\epsilon = 1$, defining our energy scale. We set our unit of length by taking $\sigma_0 = 1$. The one-body random potential $V_{d}(r)$ is constructed using a method initially devised by Chudnovsky and Dickman~\cite{chuddick}. This method generates a potential energy landscape that has zero mean and exponentially decaying short-range correlations. These take the form
\begin{equation}
 C_{V}(r)=\left[ \langle V(x) V(y) \rangle_{|x-y|=r}\right]/{\sigma}^2=\exp(-r/\xi).
 \end{equation}
The variance of the Gaussian distribution $\sigma^2$ defines the strength of the disorder. The scale set by disorder correlations is chosen to be small: $\xi = 0.12$. 

We use periodic boundary conditions across a rectangular box, defined by dimensions $L_x, L_y$ along the $x$ and $y$ directions. The number of particles $N$ is chosen so that it that accommodates a perfect triangular lattice. Our  particle densities are thus $\rho_0 = N_p/(L_x \times L_y)$. 

We compute the disorder-averaged direct pair correlation function,
\begin{equation}
 g_{r}^{(1)}(r)=\frac{1}{\rho_0^2}\left[\langle \rho(r) \rho(0) \rangle \right]-\frac{\delta(r)}{\rho_0}.
\end{equation}
The density correlation function, $g^{(2)}(r)$, (the EA-OP analog), and is defined as
\begin{equation}
 g_{r}^{(2)}(r)=\frac{1}{\rho_0^2} \left[ \langle \rho(r) \rangle  \langle\rho(0) \rangle \right].
 \label{g2r}
\end{equation}
The disorder-averaged orientational correlation functions are defined using the local hexatic order parameter,
\begin{equation}
 \psi_6^{j}=\frac{1}{n_j}\sum_{k=1}^{n_j} \exp(i6\theta_{jk}),
 \label{hex}
\end{equation}
where $n_j$ are the number of nearest neighbours of particle $j$ and $\theta_{jk}$ is the angle between the vector from particle $j$ to it's nearest neighbour $k$ and reference $x$-axis. We obtain the direct pair orientational correlations,
\begin{equation}
 g_{\theta}^{(1)}(r)=\left [ \langle \psi_6(\vec r) \psi_6^\ast (0) \rangle \right],
\end{equation}
where $\psi_{6}(\vec r)=\sum_{i=1}^{N} \delta(\vec r-\vec r_i) \psi_{6}^j$.
The orientational correlation in the Edwards-Anderson sense is defined via
\begin{equation}
 g_{\theta}^{(2)}(r)=\left[ \langle \psi_6(\vec r) \rangle \langle \psi_6^\ast (0) \rangle \right].
 \label{g2t}
\end{equation}
For simplicity we will refer to correlations that are simply the disorder-averaged correlations functions in the pure system as ``diagonal'' correlation functions, whereas the Edwards-Anderson correlation functions, non-trivial in the disordered fluid system, will be referred to as ``off-diagonal'' correlation functions. These reflect the way these correlations  are defined within a replica formulation of the statistical mechanics of disordered systems. Thus, we focus on two sets of diagonal and off-diagonal correlations, one for translational ordering and the other for orientational ordering.

The structure factor is calculated as,
\begin{equation}
 S(q)=1+\rho_0 \int \left[g^{(1)}(r)-1 \right] ~e^{i\vec q.\vec r} d \vec r,
\end{equation}
where $g^{(1)}(r)$ is the disordered radial distribution function.

To quantify the translational and orientational correlations for each density and with each disorder, we monitor the value of EA-OP $g_{r}^{(2)}(r \rightarrow 0)$, $g_{\theta}^{(2)}(r \rightarrow \infty)$, and the  first peak height of the structure factor $S(q)$.

We perform Metropolis Monte Carlo (MC) temperature quench-and-hold simulations in two dimensions keeping temperature $T = 1$. The  number of particles is varied across $N_p = 780, 1020$,and $2016$, leading to densities ranging from $\rho_0 = 0.7$ to $\rho = 1$. Disorder strengths are varied from $\sigma^2 = 0$ to $\sigma^2 = 1$, in steps of $0.1$. For a fixed $\rho$ and $\sigma^2$, we begin with an initial crystalline configuration quench the initial crystal phase to $T=1$ and hold it up to $t_{hold}=11 \times 10^5$ Monte Carlo sweeps (MCS). We do not record data for an initial $10^5$ MCS in all cases to ensure equilibration. We then average over $t^\prime=10^4$ configurations, each separated by $10^2$ MCS. Finally, we average over disorder, using  $20$ realizations for all $\sigma^2$. We will indicate thermal averages through the angular brackets $< \cdot >$ and disorder averages by $[\cdot ]$

\section{Results and Discussion}
\label{results}

\subsection{The pure case}
In this section we present the results for the correlation functions defined in Section~\ref{model} obtained from extensive MC simulations in the canonical ensemble for our system of particles interacting with the inverse twelfth power potential.  We first consider the pure system without disorder ($V_d = 0$ in Eq.~\ref{eq:hamiltonian}). Our MC simulations at $T = 1$ obtained a freezing density $\rho_0 \sim 0.986$,  in good agreement with results from earlier work~\cite{bgw}. 

\begin{figure}[ht]
\includegraphics[height=6.5cm,width=8.5cm]{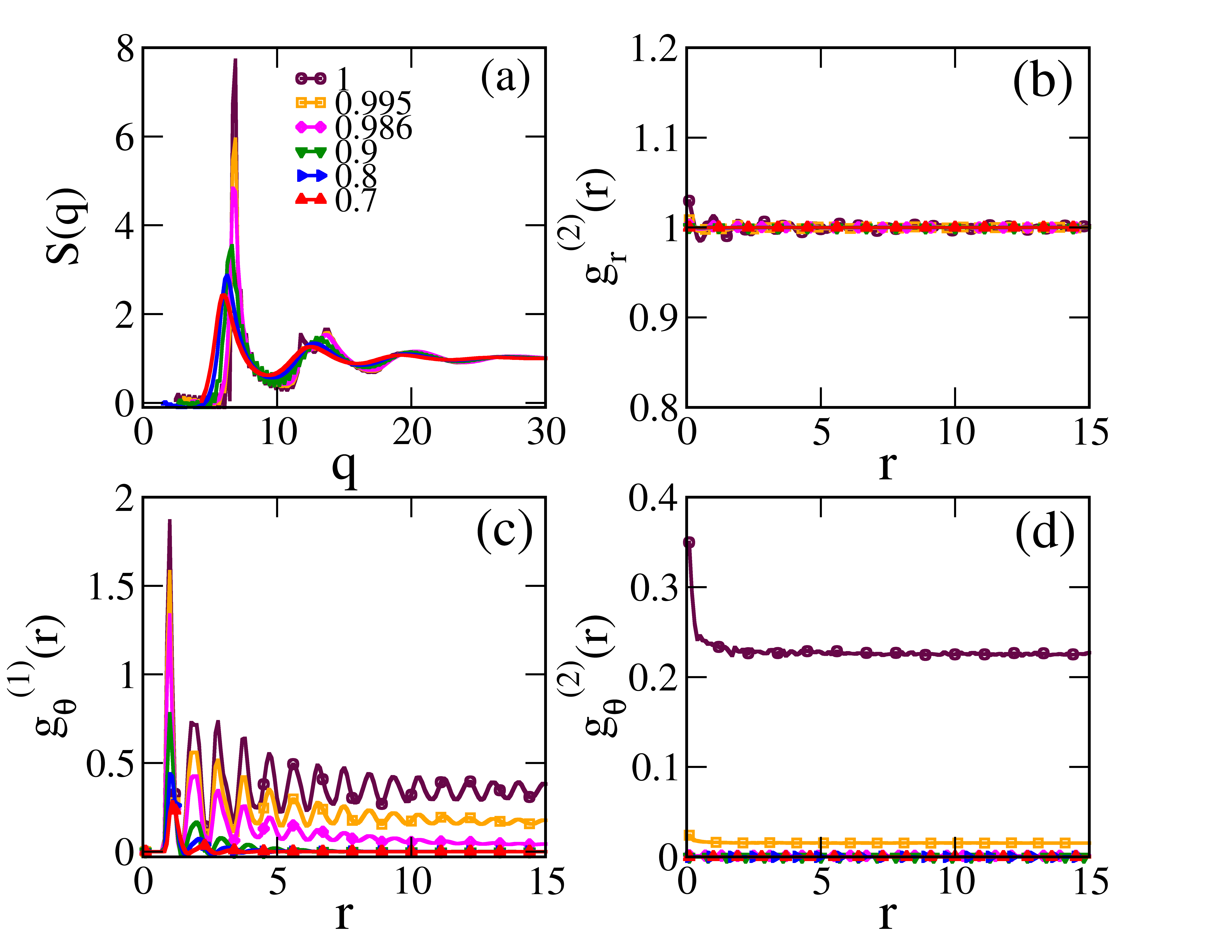}
\caption{{\it Pure system for various densities ($\rho_0$):} (a) The structure factor $S(q)$ versus $q$ and (b) the EA-OP for translations $g_{r}^{(2)}(r)$ versus $r$. The orientational correlations (c)  $g_{\theta}^{(1)}(r)$ and (d) $g_{\theta}^{(2)}(r)$ versus $r$.  Parameters: $N_p=1020$, and $\rho_0=0.7,0.8,0.9,0.986,0.995,1.$}
\label{fig1}
\end{figure}

Our first set of results, for the pure case, are  shown in Fig.~\ref{fig1}, where we present the structure factor $S(q)$, i.e. the Fourier transform of $g^{(1)}_r$ as well as $g^{(2)}_r$, $g^{(1)}_\theta$ and $g^{(2)}_\theta$, for varying densities, across Fig.~\ref{fig1}(a) - (d) . The diagonal translational and orientational correlations behave as expected (Fig.~\ref{fig1}(a)\&(c)), with correlations building up with density.  The structure factor $S(q)$ is smooth for $\rho = 0.9$, but exhibits a split second peak at $\rho_0 = 0.98$ and $0.986$. Such behavior is normally ascribed to frustration in glassy systems, but could also signify the onset of phase coexistence or the presence of long-lived local crystallinity.  

Above $\rho_0 = 1$, $S(q)$  exhibits the sharp peak characteristic of the lattice phase. the first peak height of $S(q)$ increases with density $\rho_0$. In the solid phase $g^{(1)}_\theta$ does not decay, signifying long-ranged orientational correlations. 

From Fig.~\ref{fig1}(c), we see that the orientational correlation function $g_{\theta}^{(1)}(r)$ is zero to our numerical accuracy for $\rho_0 = 0.7,0.8,0.9$, while its long-distance asymptote ${\lim}_{r \rightarrow \infty}~ g_{\theta}^{(1)}(r)$ is  finite  for $\rho_0 = 0.98,0.986, 0.99$. For larger densities, at $\rho_0=0.995, 1$, orientational correlations are unambiguously long-ranged. On the fluid side, orientational correlations at all densities are significantly longer-ranged than positional correlations. These results are consistent with prior numerical work, including results for the location of the liquid-solid transition~\cite{bgw}.

The off diagonal correlations on the other hand are expected to be trivial; see (Fig.~\ref{fig1}(b)\&(d)). The EA-OP for translations is uniform at $g_{r}^{(2)}=1$ for $\rho_0 = 0.7,0.8,0.9, 0.98$ and $0.986, 0.99$.  There are very small and negligible fluctuations for larger densities $\rho_0=0.995, 1$, likely originating from long-lived hexatic~\cite{KT,kthny1,kthny2,kthny3} or crystal fluctuation. The EA-OP for orientations is zero for $\rho_0 = 0.7,0.8,0.9$. From Fig.~\ref{fig1}(d), we see that closer to the transition, it decays sharply to zero, for $\rho_0 = 0.98,0.986,0.99$, with residual structure at short-scales possibly associated with either long-lived hexatic~\cite{KT,kthny1,kthny2,kthny3}  correlations or a solid phase (see Section~\ref{conclude} for further details). The EA-OP decays to a finite value for larger densities $\rho_0 = 0.995$~\& $1$. 

These results, for the pure system, are also consistent with a transition between liquid and crystalline phases, in the presence of a narrow coexistence regime. Our simulations do not attempt to resolve the possibility of two-stage melting or more complex scenarios, rather our interest is largely in disorder-induced correlations in the well-defined fluid phase.

\begin{figure*}[ht]
\begin{center}
\includegraphics[height=18cm, width=15cm]{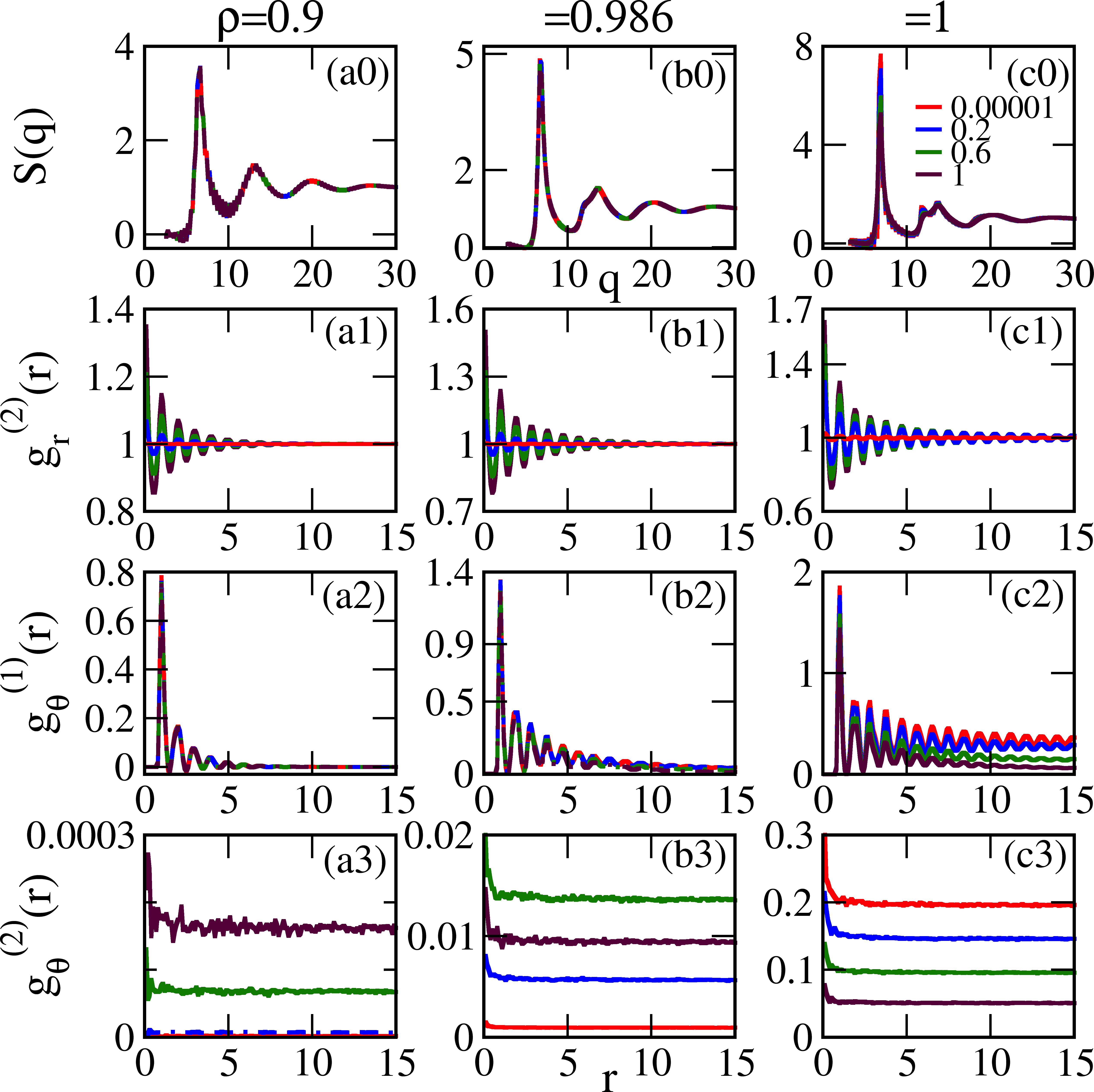}
\caption{{\it The structure factor, translational, orientational correlations with disorder:} The structure factor $S(q)$ vs $q$ (a0,b0,c0), the EA-OP for translations $g^{(2)}(r)$ vs $r$ (a1,b1,c1), the orientational correlations $g_{\theta}^{(1)}(r)$ vs $r$ (a2,b2,c2), and $g_{\theta}^{(2)}(r)$ vs $r$ (a3,b3,c3) for densities $\rho=0.9,0.986,1$ respectively. Simulation parameters: $N_p=1020$, and $\sigma^2=0.00001,0.2,0.6,1$.}
\label{fig2}
\end{center}
\end{figure*}
\begin{figure*}[ht]
\begin{center}
\includegraphics[height=10.0cm, width=16cm]{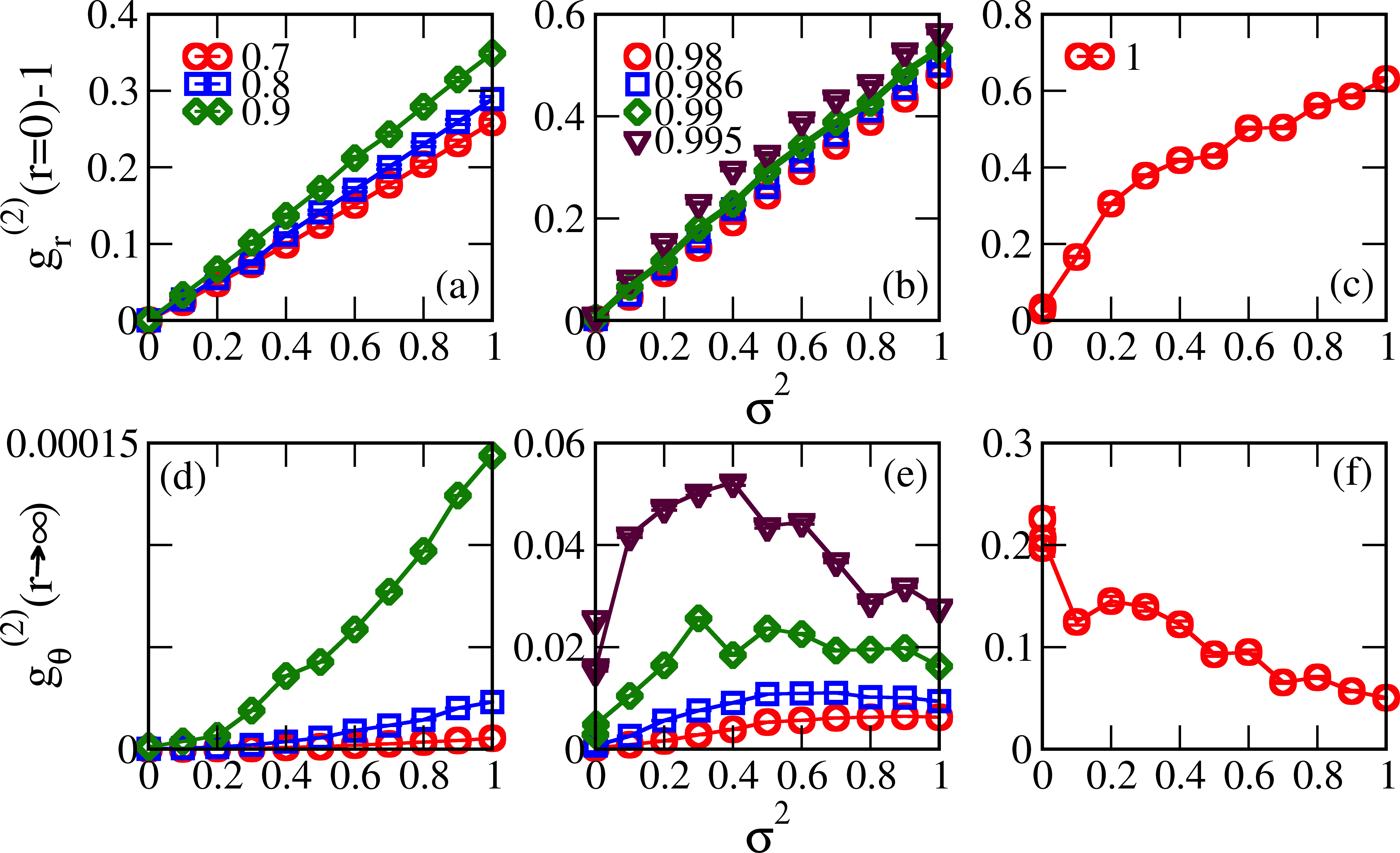}
\caption{{\it The EA-OP correlations in translation and orientation with disorder for different densities:} The EA-OP correlations in short-distance translations ($g_{r}^{(2)}(r=0)-1$) versus disorder ($\sigma^2$) and long-distance orientations ($g_{\theta}^{(2)}(r \rightarrow \infty)$) versus disorder ($\sigma^2$) for densities $\rho_0=0.7,0.8,0.9$ (a,d), $\rho_0=0.98,0.986,0.99,0.995$ (b,e), and $\rho_0=1$ (c,f).}
\label{fig3}
\end{center}
\end{figure*}
\begin{figure*}[ht]
\begin{center}
\includegraphics[height=5.0cm, width=16cm]{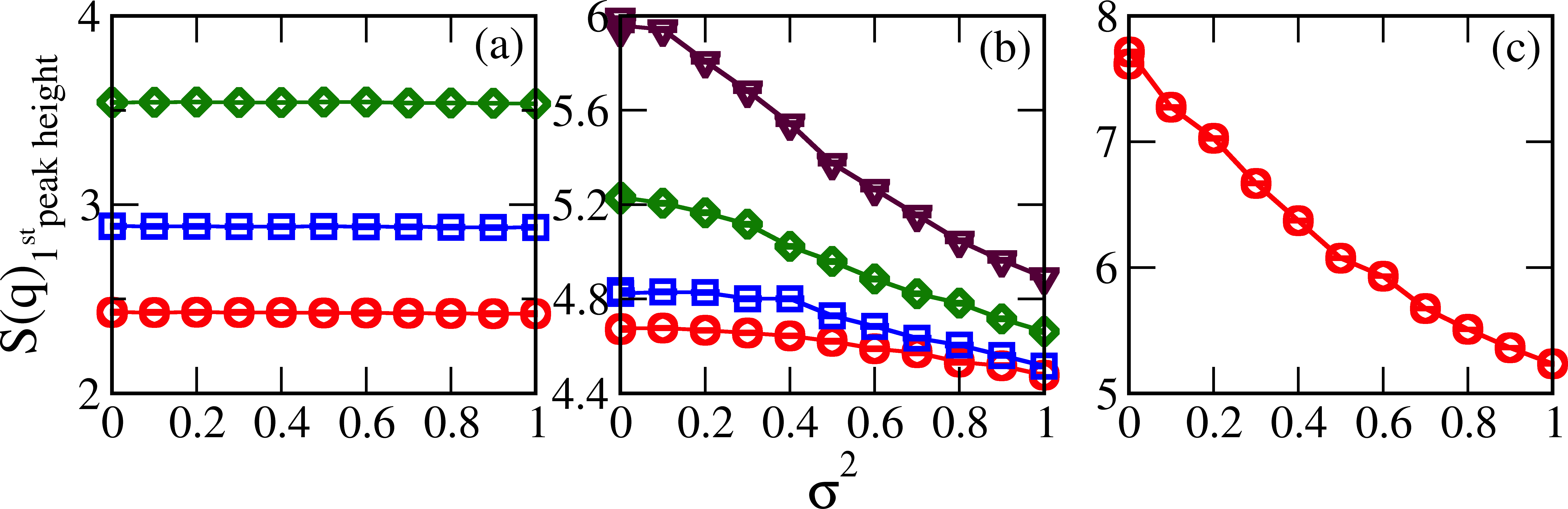}
\caption{{\it The structure factor with disorder:} The first peak height of the structure factor $S(q)_{1^{st} peak height}$ vs disorder $\sigma^2$ for densities $\rho_0=0.7,0.8,0.9$ (a), $\rho_0=0.98,0.986,0.99,0.995$ (b), and $\rho_0=1$ (c) respectively.}
\label{fig4}
\end{center}
\end{figure*}

\subsection{The disordered case}
We now consider the effect of a non zero random potential ($V_d \neq 0$ in Eq.~\ref{eq:hamiltonian}). We vary particle densities between $\rho_0 = 0.7$ and $\rho_0 = 1$,  and disorder strengths  from $\sigma^2=0.00001$ to $\sigma^2 = 1$. Our simulations thus cover 5 orders of magnitude in disorder strength. 

Our results are presented in Fig.~\ref{fig2}.  For $\rho_0 = 0.9$, the structure factor and the other correlation functions are shown in Fig.~\ref{fig2} (a0-3). The structure factor, essentially the Fourier transform of $g_{r}^{(1)}(r)$ has smooth peaks, as seen in the pure system. Its dependence  on $\sigma^2$ is negligible to first approximation. The EA-OP for translations decays rapidly quickly to $g_{r}^{(2)}(r)=1$, although its oscillations are amplified as $\sigma^2$. The diagonal correlation function $g_{\theta}^{(1)}(r)$ decays exponentially  to zero as in the liquid phase of pure system. Its behaviour seems to be independent of disorder strength at this relatively low density value. The off-diagonal orientational correlations, monitored through $g_{\theta}^{(2)}(r)$, are numerically small and fluctuate asymptotically around a small non-zero value that  increases with $\sigma^2$.

Fig.~\ref{fig2} (b0-3) shows orientational and translational correlations for $\rho_0 = 0.986$, across a range of disorder strengths. Again the structure factor is unaffected by disorder. The $g_{r}^{(2)}(r)$ decays to unity in an oscillatory manner, and the oscillations are amplified  as $\sigma^2$ increases. The $g_{\theta}^{(1)}(r)$ now shows a weak dependence with $\sigma^2$. The $g_{\theta}^{(2)}(r)$ remains numerically small, decaying to a non-zero value.  

Finally, for $\rho_0 = 1$, where a crystal is expected in the pure system, we show diagonal and off-diagonal correlations in Fig.~\ref{fig2} (c0-3). The structure factor $S(q)$ has a split in the second peak. The EA-OP for translations decays to unity at large $r$ and the oscillations are amplified with $\sigma^2$. The $g_{\theta}^{(1)}(r)$ now exhibits ordering that persists across the size of the simulation box.  The $g_{\theta}^{(2)}(r)$ oscillates prominently while decaying to its asymptotic value. 

In Figs.~\ref{fig3} and~\ref{fig4}, we summarize our results for the asymptotic values of $g_{r}^{(2)} (r=0)$, $g_{\theta}^{(2)}(r \to \infty)$ as well as the structure factor $S(q)$, across a range of densities and disorder strengths. Note that in a finite size simulation box with periodic boundary conditions, the largest distance that is possible is $L/2$, where $L$ is the length of the simulation box, The most sensitive probes of the effects of both disorder and interactions comes from considering $g_{r}^{(2)}(r=0)$ and $g_{\theta}^{(2)}(r \rightarrow \infty)$. In addition, as a probe of the effects of disorder on the conventionally measured two point positional correlation functions, we also show  $S(q = q_{\rm max})$, the magnitude of the first peak of the structure factor,  for a range of densities $\rho$ and disorder strengths $\sigma^2$.

\subsection{Reentrant short-ranged order  in $g_\theta^{(2)}(r)$}

We exhibit results for varying $\rho_0$ across three ranges of values: $0.7$, $0.8$ and $0.9$ in Figs.~\ref{fig3} (a)\&(d); $0.98$, $0.986$, $0.99$ and $0.995$ in FIgs.~\ref{fig3} (b)\&(e); and $1$ in Figs.~\ref{fig3}(c)\&(f). We find that for the lowest densities $g^{(2)}(r =0 )$ and $g_{r}^{(2)}(r \rightarrow \infty)$ both increase with $\sigma^2$. Close to the boundary between liquid and solid but within the putative liquid phase, we examine these correlation functions as shown in Figs.~\ref{fig3} (b)\&(e). The  value of  $g^{(2)}(r = 0)$ increases linearly with $\sigma^2$ while the $g_{r}^{(2)}(r \rightarrow \infty)$ value appears to show a {\it re-entrant behavior} with increase in disorder strength. We also present results for the solid regime, for completeness, at the single density value of  $\rho=1$ as shown in Fig. 3 (c)\&(f). The value of $g^{(2)}(r=0)$ increases with $\sigma^2$  while, surprisingly, the value of $g_{r}^{(2)}(r \rightarrow \infty)$ decreases with $\sigma^2$. 
 
The height of the first peak of the structure factor, $S(q = q_{max})$ remains roughly constant with $\sigma^2$ as shown in Fig.~\ref{fig4} (a). This indicates that deep into the fluid phase, conventional disorder-averaged correlation functions are only very weakly affected by the disorder. The Edwards-Anderson correlation function, in contrast, disperses more strongly both with disorder and with interactions.The height of the first peak of the structure factor $S(q)$ is roughly independent of disorder for smaller $\sigma^2$, while it decreases linearly for larger $\sigma^2$ as shown in Fig.~\ref{fig4} (b). Finally the same quantity  decreases with $\sigma^2$ as shown in Fig. ~\ref{fig4} (c) showing that disorder, as expected, destroys translational order.

As we have discussed, orientation correlations, at the level of two point functions, can be characterized in two different ways. The diagonal correlator, $g_\theta^{(1)}(r)$, measures equal-time correlations of the product of densities within the same disorder background, with a subsequent averaging over disorder realizations performed to ensure that this correlation function regains translational invariance. In contrast, the off-diagonal correlator $g_\theta^{(2)}(r)$ measures correlations between \emph{time-averaged densities}, which is then averaged over realizations of the quenched disorder field. This is trivial in the absence of disorder, since translational invariance implies that the local density at any point in a fluid is the same. 

However in a disordered system, this quantity is more subtle. Its main contribution comes from the pinning of two particles at separated points. While, in a fluid, orientational order about any one single pinning site separately can vary continuously, the presence of two such points defines an axis in the fluid that tends to ``lock-in'' orientational order at these points. This is the origin of non-trivial structure in $g_\theta^{(2)}(r)$. As the disorder potential is increased, so too is the tendency to such lock-in accentuated, as particles spend more time pinned to locations that are deeper in the potential landscape. This provides an explanation for why $g_\theta^{(2)}(r)$ at intermediate ranges, should increase with increasing disorder, as seen in the data here where, for the liquid regime at densities $\rho_0=0.9,0.8,0.7$.

Why does the $g_\theta^{(2)}(r)$ at intermediate ranges \emph{decrease} with disorder in the solid phase? We believe that this is most easily linked to the proposal that the low-temperature state of a disordered solid in two and three dimensions most resembles a domain arrangement~\cite{gauflux,gauflux2}.  Within each domain, orientational and translational order is maintained, but de-correlates rapidly at larger scales.The characteristic domain size decreases rapidly as disorder increases.This suggests that  $g_\theta^{(2)}(r)$ at intermediate ranges should be controlled by the rapid collapse of domain sizes with increasing disorder in the solid, consistent with this picture. Close to the boundary separating solid from liquid, the re-entrant phase behaviour can be understood in terms of the opposing behaviour in the liquid and solid phases detailed above.

\begin{figure*}[ht]
\begin{center}
\includegraphics[height=10cm, width=16cm]{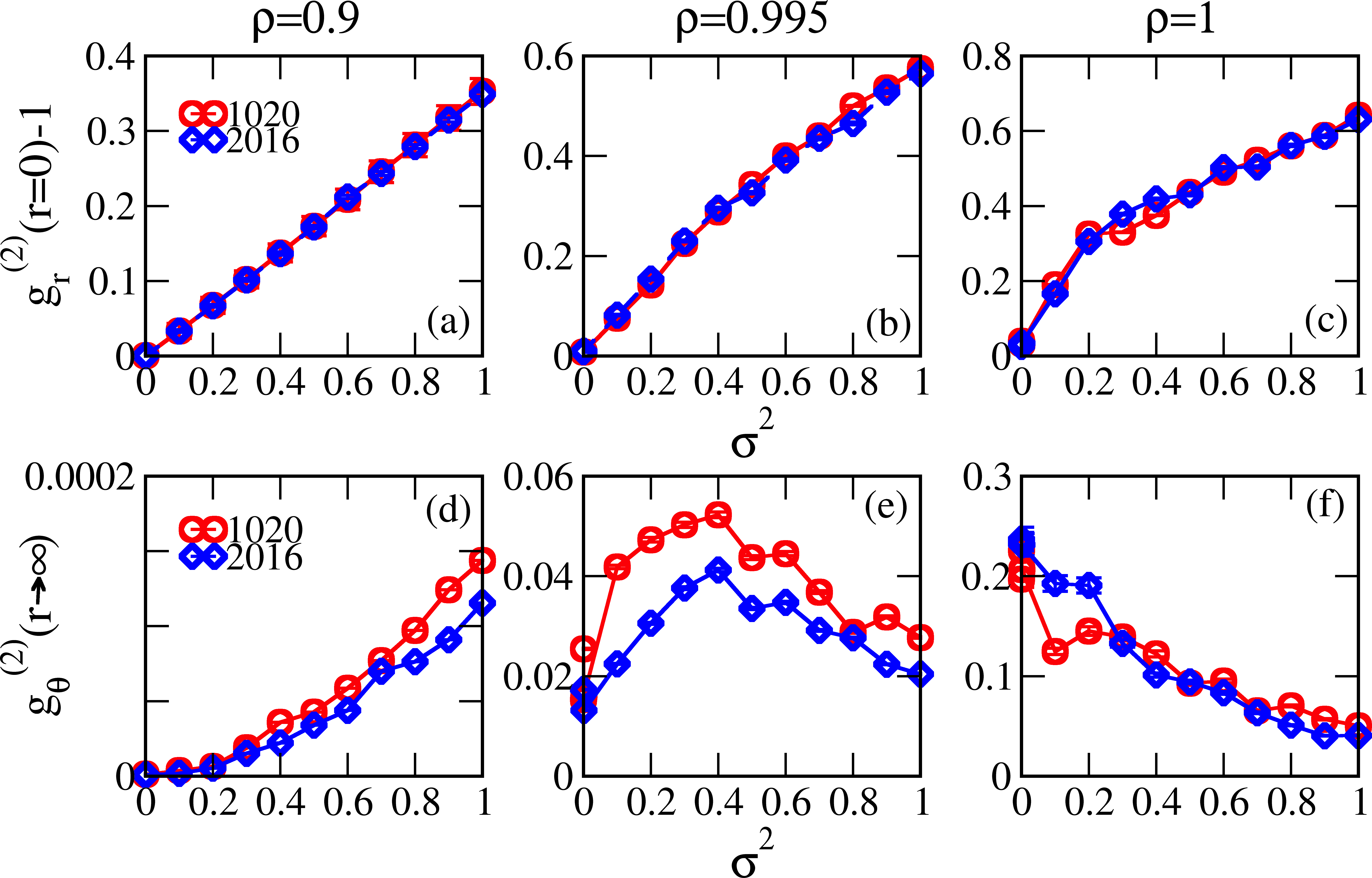}
\caption{{\it The finite size effect of EA-OP correlations in translation and orientation with disorder for different densities:} For system sizes $N_p=1020, 2016$, the EA-OP correlations in short-distance translations ($g_{r}^{(2)}(r=0)-1$) versus disorder ($\sigma^2$) and long-distance orientations ($g_{\theta}^{(2)}(r \rightarrow \infty)$) versus disorder ($\sigma^2$) for densities $\rho_0=0.9$ (a,d), $\rho_0=0.995$ (b,e), and $\rho_0=1$ (c,f) respectively.}
\label{fig5}
\end{center}
\end{figure*}

 \subsection{Finite size effects}

We further examined the effects of the finite size of the simulated system. The effects of a finite system size on $g_{r}^{(2)}(r=0)$, $g_{\theta}^{(2)}(r \rightarrow \infty)$ are shown in Fig.~\ref{fig5}. It is clear that $g_{r}^{(2)}(r=0)$ linearly increases with $\sigma^2$ and that the sensitivity to system size is relatively weak. The value of $g_{\theta}^{(2)}(r \rightarrow \infty)$ decreases with the system size $N_p$ in all three regimes, although the phenomenon of {\it re-entrance} is insensitive to the system size $N_p$. This is most prominent at densities in the vicinity of the liquid-solid transition in the pure system. Given that we must also average over disorder for our results to be meaningful, our systems do not span a sufficiently large range in $N$ for a systematic finite-size scaling analysis to be possible. Thus, we leave open the possibility that  $g_{\theta}^{(2)}(r \rightarrow \infty)$ vanishes in the true asymptotic limit of $N \rightarrow \infty, r \rightarrow \infty$. (As pointed out earlier, in a finite size simulation box with periodic boundary conditions, the largest distance that is possible is $L/2$, where $L$ is the length of the simulation box and it is this scale that we actually probe.) However, we believe that our results for order at intermediate ranges, of about 10 - 30 inter-particle spacings should be robust. It is these scales that can be probed most easily in experiments, given that the field that can be imaged must contain a limited number of particles at any time.

In our system, $g_\theta^{(1)}(r)$ {\em always} vanishes in a liquid at large $r$ while in a solid, it attains a constant value indicating simply that the solid phase is both positionally and orientationally ordered. An intermediate hexatic phase has never been observed during melting of the inverse twelfth power crystal~\cite{bgw}.  We find, in this paper, that off-diagonal orientational correlations increase with the strength of quenched disorder in the range of densities where the pure system is a liquid. For higher densities, $g_\theta^{(2)}(r)$ for large $r$, decreases with disorder. These two results, taken together, suggests that quenched disorder, enhances orientational correlations in a liquid but destroys orientational order in a crystal. Surprisingly, in an intermediate density range, there is a re-entrant behavior, where $g_\theta^{(2)}(r)$ at large $r$ first increases and then decreases as disorder is increased. These results are quite unusual because $g_\theta^{(2)}(r)$ is defined (see Eq.~\ref{g2t} in Section~\ref{model}) such that it should vanish if the orientational order parameter $[\langle \psi_6(\vec r) \rangle]$ itself vanishes. Better statistics by simulation over larger times and larger system sizes do not affect our qualitative results significantly. We attribute this to a ``lock-in'' mechanism aligning local orientational correlations long the axes joining well-pinned particles, as explained above, possibly also aided by the presence of a metastable or even stable hexatic phase. 

\section{Summary and Conclusions}
\label{conclude}
In this paper, we presented calculations of disorder-averaged orientational and translational correlations, varying densities and disorder strengths, in a relatively simple model system. We considered a collection of particles in two dimensions interacting with the inverse twelfth power potential~\cite{bgw}, and under the influence of a spatially varying, quenched, random field with short-ranged correlations~\cite{gauchand,ankeur,egelme}. We introduced a new orientational correlation function,  with non-trivial values only in systems of interacting particles with quenched disorder. The corresponding quantity involving positional correlations has been studied in the past using liquid state theory~\cite{gauchand}, simulations~\cite{ankeur} and experiments~\cite{egelme}. These correlation functions, both positional and orientational, may be computed in a straightforward fashion for configurations of particles obtained either in simulations or experiments. 

Our interest in computing these was connected to the possibility of measuring them directly in the fluid phase of a colloidal system, where disorder can be introduced using spatially random, static light fields~\cite{egelme,jorg-rev}. Such experiments can be used to generate a large number of snapshots of particles, from which any equal-time correlation can be computed, including the ones we describe here. The disorder averaging can be done either by considering a sufficiently large system, assuming self-averaging, or by equilibrating configurations in a variety of underlying disorder configurations with the same statistical properties.

The quantity $g_\theta^{(1)}(r)$ in a pure system, where the requirement of a separate averaging over disorder realizations is absent, has been studied before in the context of the problem of dislocation mediated two dimensional melting~\cite{CL, KT, kthny1,kthny2,kthny3} of a disorder free solid. It has been used to detect the possible presence of a {\it hexatic} phase. In a hexatic, while positional correlations decay exponentially with distance, $g_\theta^{(1)}(r)$ decays very slowly, as a power law~\cite{kthny1}. This has to be contrasted with a liquid, where both positional and orientational correlations both decay exponentially. 

The current understanding of the phase diagram of two-dimensional melting can be summarized as follows. While it appears that two-stage melting, as proposed in the Berezinskii, Kosterlitz, Thouless, Halperin, Nelson, Young (BKTHNY) theory~\cite{kthny1,kthny2,kthny3} is the norm in sufficiently long, well-equilibrated simulations, the two separate transitions (between crystal and hexatic and then between hexatic and liquid) need not both be continuous, as the theory originally suggested. The BKTHNY theory is a renormalization group approach~\cite{CL} that assumes a high defect core energy a priori to construct a perturbation expansion. However, defect core energies are properties of the underlying interaction potential as well as the density, and one could anticipate the possibility of both continuous and discontinuous transitions in general, as well as the possibility that both transitions might be coincident.  Bernard and Krauth~\cite{BernardKrauth,Engel,Qi} have suggested that the  transition from crystal to hexatic is generically continuous one, whereas  the transition from hexatic to isotropic liquid is first-order. Alternate scenarios involving a first-order crystal to liquid melting has also been discussed~\cite{Chui,TVR}. 

A series of papers by Ryzhov and collaborators ~\cite{ryzhov1,ryzhov2,ryzhov3,ryzhov4,ryzhov5,ryzhov6} use core-energy-based calculations to  show that accounting for the core energies of dislocations and disclinations can yield  different scenarios for two-stage melting, including both first-order and continuous transitions. In particular it has been suggested that for repulsive potentials $1/r^n$ with $n > 6$, the Bernard-Krauth scenario holds whereas for $n < 6$, the older scenario of two continuous transitions is standard. In all cases, the intervening hexatic regime is very narrow~\cite{kapfer, amir}.  An equilibrium hexatic has been detected for many soft colloids~\cite{col2d} such as hard discs~\cite{jaster}, super-paramagnetic colloidal systems~\cite{maret1,maret2} and Gaussian core models~\cite{gaucore}. Whenever such a phase occurs, it is observed to lie very close to the liquid-solid boundary and is, in principle, difficult to distinguish from the usual liquid-crystal two phase coexistence~\cite{mywork}. 

Finally, quenched disorder is known to enhance the stability of the hexatic phase~\cite{maret3}. In amorphous or model glassy solids, local analysis of $g_\theta^{(1)}(r)$ often detects so called ``medium range crystalline order'', which may not be distinct from a hexatic induced by (annealed) disorder in the interactions~\cite{tah,tanaka}.  

The central result of this paper is that quenched disorder enhances medium-range orientational correlations in a liquid but destroys it in a crystal. In a range of densities intermediate between that of the solid and of the liquid, we see a re-entrant behavior, where $g_\theta^{(2)}(r)$ at large $r$ first increases and then decreases as disorder is increased. Our results, unfortunately cannot distinguish between whether a true hexatic is stabilized by quenched disorder through pinning or pinning increases the proportion of the ordered solid phase in a thin, liquid-solid coexistence region. Distinguishing between these scenarios will require much more analytic as well as computational efforts beyond the scope of this paper. We point out that experiments on two-dimensional colloidal fluids in random potentials can be performed and that studies of the fluid phase are far less susceptible to issues of metastability than the disordered solid. 

Probing correlations and response in the disordered liquid may shed light on the reentrant behaviour we describe here. They should also provide benchmarks for the unusual Edwards-Anderson correlation function whose properties are defined and studied in this paper. 

\begin{acknowledgments}
 The authors acknowledge many useful discussions with Stefan Egelhaaf. We thank Pinaki Chaudhuri and Chandan Dasgupta for a number of useful conversations. Financial support from the DAE, Gov. of India is gratefully acknowledged. This work was partially supported by the PRISM project at IMSc (GIM) and the Shastri Mobility Program of the Shastri Indo-Canadian Institute (GIM). 
 
 \end{acknowledgments}

\end{document}